\documentclass{moriond}

\usepackage{graphicx}

\graphicspath{{./fig/}}

\usepackage{amsmath}

\def\be{\begin{equation}}
\def\ee{\end{equation}}
\def\bea{\begin{eqnarray}}
\def\eea{\end{eqnarray}}
\usepackage{amsmath}

\newcommand{\Photo}{\includegraphics[height=35mm]{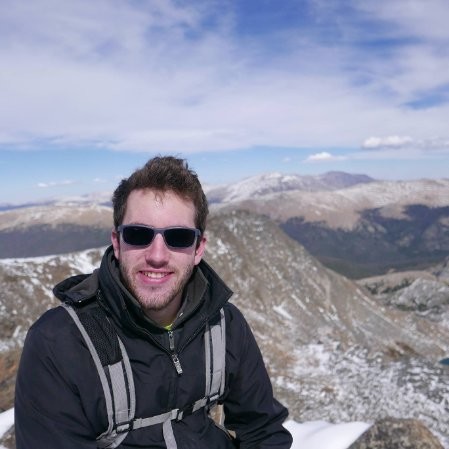}}

\begin{document}
\vspace*{4cm}
\title{Gravitational radiation from MHD turbulence in the early universe}

\author{ A. Roper Pol}

\address{Universit\'e de Paris, CNRS, Astroparticule et Cosmologie, Paris, F-75013, France\\
School of Natural Sciences and Medicine, Ilia State University, 3-5 Cholokashvili Street, 0194 Tbilisi, Georgia}

\maketitle
\abstracts{
I briefly discuss recent results of numerical simulations addressing the generation of a
cosmological gravitational wave background produced by turbulence sources in the early
universe.
Contribution to the 2021 Gravitation session of the 55th Rencontres de Moriond.}

\section{Introduction}

The field of gravitational astronomy is in a period of blossoming.
Starting in 2015 with the first detected gravitational wave event GW150914, the LIGO-Virgo collaboration has recently released the first half of the third observing run O3a, with 39 new
observed events, amounting to a total of 50 detected events.\cite{Abbott:2020niy}
In the 1--100 nHz regime, pulsar-timing arrays (PTA) seek to detect GWs by monitoring the
time-delay signals of an array of millisecond pulsars.
Recently, it has been reported by the North American Nanohertz Observatory for
Gravitational Waves (NANOGrav) after 12.5 years of observation a common-spectrum process, 
although the statistical significance of a quadrupolar correlation, as expected for GW signals,
is still not conclusive.\cite{Arzoumanian:2020vkk}
Even though most of the current applications focus on astrophysical events, cosmological
applications are becoming more relevant, and space-based GW detectors, such as the Laser
Interferometer Space Antenna (LISA), PTA and CMB anisotropies might even allow us to
explore the first moments of the history of our Universe.
Many sources of GWs in the early universe have been proposed, and how to
disentangle the different contributions is a complicated problem that requires a lot of work from
the GW community in the years to come.
An interesting source of GWs in the early universe corresponds to turbulence,
which can be generated by violent events, such as first-order cosmological phase transitions
or by primordial magnetic fields due to the high conductivity of the primordial plasma producing
magnetohydrodynamic (MHD) turbulence.
Previous analytic estimates have suggested that turbulence generated at the electroweak
phase transition (EWPT) can produce GWs that can be detected by LISA.\cite{%
Kosowsky:2001xp,Caprini:2019egz}
Although semi-analytical models of MHD turbulence have been used to predict the GW spectral
shape,\cite{Niksa:2018ofa}
the highly non-linear dynamical evolution of MHD turbulence requires the use of numerical 
simulations.
I present an overview of the status of such simulations and a discussion of some of the latest
relevant results.

\section{Magnetohydrodynamic turbulence and GW production}

The stress tensor that leads to GW production is sourced by velocity and magnetic fields,
\begin{equation}
T_{ij}=(p/c^2 + \rho) \gamma^2 u_i u_j - B_i B_j + \delta_{ij} B^2/2,
\label{eq:Tmunu}
\end{equation}
where we have neglected electric fields due to the high conductivity of the early universe
during the radiation-dominated era.
The magnetohydrodynamic (MHD) conservation laws on an isotropic and homogeneous
universe described by the Friedmann-Lema\^itre-Robertson-Walker (FLRW) metric tensor,
with a radiation equation of state $p=\rho/3$, are
\begin{align}
\partial_t \ln \rho = &\, - \frac{4}{3} \left(\nabla_j u_j + u_j \nabla_j \ln \rho \right) + \frac{1}{\rho} \left(
u_j \varepsilon_{jmn} J_m B_n + \eta J^2 \right), \\
D_t u_i = &\, \frac{1}{3} u_i \left(\nabla_j u_j + u_j \nabla_j \ln \rho \right) - \frac{u_i}{\rho} 
\left(u_j \varepsilon_{jmn} J_m B_n + \eta J^2 \right) \nonumber \\
 - &\, \frac{1}{4} \nabla_i \ln \rho + \frac{3}{4\rho} \varepsilon_{imn} J_m B_n + \frac{2}{\rho} 
\nabla_j \rho \nu S_{ij} + {\cal F}_i, \\
\partial_t B_i = &\, \varepsilon_{imn} \nabla_m \left(\varepsilon_{npq} u_p B_q - \eta J_n +
{\cal E}_n \right),
\label{eq:MHD}
\end{align}
where the magnetic and density fields are expressed in units normalized by the radiation energy
density at the time of turbulence generation, and are comoving with the Universe expansion, and the time is normalised to the initial conformal time of turbulence generation.\cite{%
Brandenburg:1996fc,Brandenburg:2017neh,Pol:2018pao,Pol:2019yex}
We consider two main sources of turbulence in the early universe:
\begin{enumerate}
\item  Hydrodynamic turbulence induced by a first-order phase transition.
\item Primordial magnetic fields that are produced or present at or during a phase transition.
\end{enumerate}
The turbulence is characterized by the spectral peak $k_*$ of the sourcing (velocity or magnetic) field.
The scale $k_*$ of the turbulence is comoving and normalized by the Hubble scale $H_*$ at the time of generation.
Physically motivated values of $k_*$ are about 100 for the EWPT\,\cite{Turner:1992tz} and
$10$ for the QCD phase transition.\cite{Witten:1984rs}
We consider the initial energy density of the turbulent source to be a fraction of the radiation energy density at the time of generation that
cannot exceed 10\% due to the Big Bang nucleosynthesis limit.\cite{Kahniashvili:2009qi}
In the case of magnetic fields, Fermi observations of blazar spectra give lower limits on the
magnetic field amplitudes present in cosmic voids.\cite{Neronov:1900zz}
Assuming that these fields have evolved from cosmological seeds allows us to impose lower bounds on primordial
magnetic fields that have been generated at some early cosmological epoch.
The presence of partial helicity seems to be required to explain the observed lower bounds at
very large scales if the magnetic field is produced at the electroweak scale.\cite{Brandenburg:2017neh}

The tensor-mode perturbations $h_{ij}^{\rm phys}$ above the FLRW background
are described by the GW equation. For anisotropic sources, it reads\,\cite{Grishchuk:1974ny,Deryagin:1986qq}
\begin{gather}
\left(\partial_t^2 - \nabla^2\right) h_{ij} = 6T_{ij}^{\rm TT}/t,
\label{GW_eq_norm}
\end{gather}
for scaled strains $h_{ij}=ah_{ij}^{\rm phys}$, comoving coordinates and stress tensor components,
and conformal time.
We have used a normalization appropriate for numerical simulations,\cite{Pol:2018pao,Pol:2019yex}
and TT corresponds to the traceless-transverse projection.

\section{Numerical results}

The set of partial differential equations given in the previous section can be computed by performing
numerical simulations, in which we solve the MHD equations for the
stress tensor, $T_{ij}$, and then compute the resulting GW radiation, sourced by its TT projection.
The open-source {\sc Pencil Code}\,\cite{Brandenburg:2020oxc}
contains a module for GW computations that has been recently added,\cite{Pol:2018pao}
and that is being used in recent studies of GWs sourced by early-universe
turbulence.\cite{Pol:2019yex,Kahniashvili:2020jgm,Brandenburg:2021aln,Brandenburg:2021bvg,Brandenburg:2021tmp,He:2021bqm}
Different turbulent scenarios can be considered to model the initial
magnetic or velocity fields.\cite{Pol:2019yex}
\begin{enumerate}
\item Turbulent primordial magnetic fields present at the phase transition with a characteristic scale given as a fraction of the Hubble scale.
When the EWPT is considered, the resulting spectrum peaks around the maximum LISA
sensitivity when $k_* \sim 2\pi \times 100$, and we reach PTA sensitivities when the
characteristic scale is comparable to the horizon of the QCD phase transition.
\item Magnetic or velocity fields sourced during a short period of time ($\sim 0.1 H_*^{-1}$)
via a forcing term introduced in the induction or the momentum equations, respectively.
The resulting GW production is enhanced,\cite{Pol:2019yex,Kahniashvili:2020jgm}
and hence the prospects of detectability are improved.
The results show that acoustic turbulence (from velocity fields) is more efficient in producing
GW than vortical turbulence (both from magnetic and velocity fields),\cite{Pol:2019yex} as
it had been previously computed numerically for sound waves.\cite{Hindmarsh:2015qta}
\end{enumerate}
In both cases, the GW is generated in a very short amount of time ($\delta t \sim 1/k_*$)
after the energy of the source has reached its maximum value.
The time evolution of the small wave number modes of GW energy density shows a time increase proportional
to $\delta t^2$ at earlier times.
Following such time evolution of the different modes, the subinertial range of the GW spectrum (with
spectral peak
$k_{\rm GW} = 2k_*$, due to being sourced by the convoluted magnetic and/or velocity fields)
presents $\Omega_{\rm GW}\sim k$ spectrum below the peak down to scales of the order of the
Hubble scale.
Previous analytical estimates predicted
$\Omega_{\rm GW}\sim k^3$ spectrum below the peak, so this is a novel result of the
numerical simulations.\cite{Pol:2019yex}
Slightly different power laws can be observed depending on the specific dynamics\,\cite{Brandenburg:2021aln,Brandenburg:2021bvg,Brandenburg:2021tmp,He:2021bqm}
and the specific description is still on-going work.
Longer times of forcing have been considered,\cite{Kahniashvili:2020jgm} as well as magnetic
fields produced by the chiral magnetic effect.\cite{Brandenburg:2021tmp}
The effects of non-Gaussian turbulent fields (potentially produced by the dynamical MHD evolution) can
also alter the slopes of the stress and the resulting GW spectra.\cite{Brandenburg:2021bvg,Brandenburg:2019uzj}
The $k^3$ power law is observed to appear at short times\,\cite{Pol:2019yex}
while the GW amplitude is growing
and it is expected to be kept at super-horizon scales due to causality.
However, simulations that include super-horizon scales and characteristic scales require a
very large dynamical range and are challenging.
In general, vortical sources with a total energy density of 1-10\% of the radiation energy
can source GW signals detectable by LISA with a signal-to-noise ratio (SNR) of 10.\cite{Pol:2019yex}

\begin{figure}
\begin{minipage}{0.5\linewidth}
\centerline{\includegraphics[width=\linewidth]{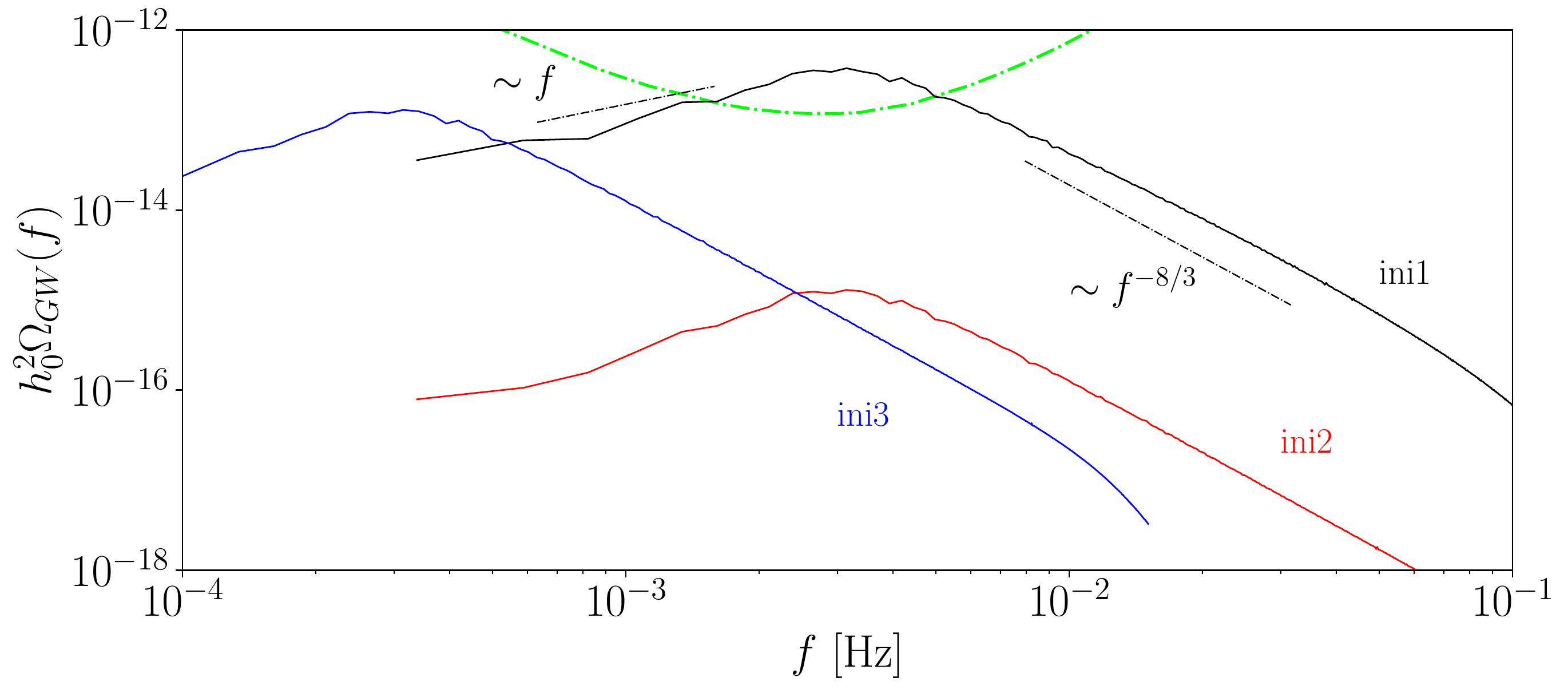}}
\end{minipage}
\hfill
\begin{minipage}{0.5\linewidth}
\centerline{\includegraphics[width=\linewidth]{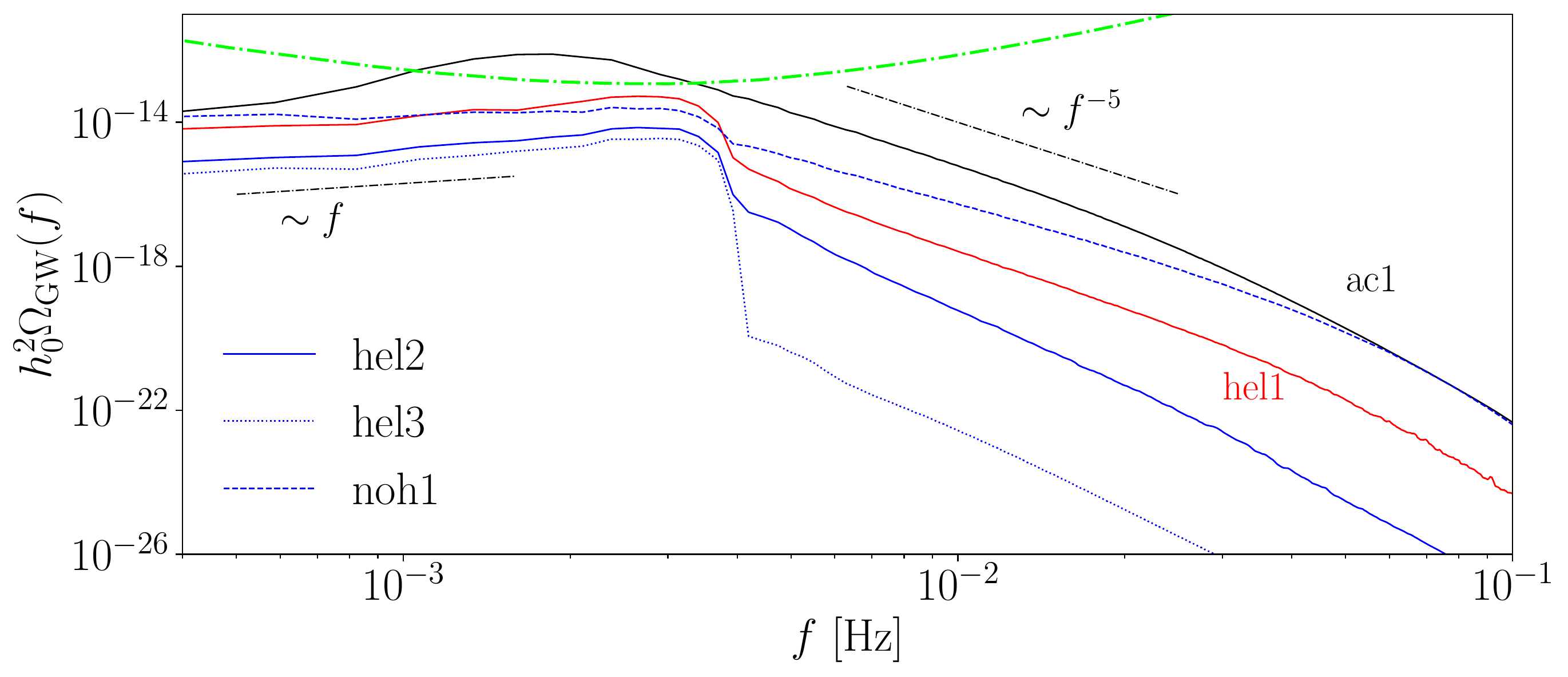}}
\end{minipage}
\hfill
\caption[]{GW spectrum for initial turbulent magnetic fields (left) and for velocity and magnetic turbulent fields produced by shortly forcing the MHD equations (right).
The GW signal is assumed to be generated at the electroweak phase transition, and compared to
the power law sensitivity of LISA using a SNR of 10 and 4 years of mission duration.\cite{Caprini:2019pxz}
The exact parameters of each run can be found in Roper Pol et al. 2021.\cite{Pol:2019yex}}
\label{fig:radish}
\end{figure}

Parity-odd violating processes during the early universe lead to circularly polarized
GW signals.
The degree of circular polarization can be computed from the results of the numerical
simulations and it has recently been studied numerically in the case of stationary turbulence\cite{Kahniashvili:2020jgm}
The detection of polarization using the dipolar response of LISA
induced by our proper motion\,\cite{Domcke:2019zls} or combining LISA and other
space-based GW detectors, e.g., TianQin,\cite{Orlando:2020oko} from different
turbulent sources using numerical simulations is on-going work by the author and
collaborators.

Following the recently reported detections by NANOGrav,\cite{Arzoumanian:2020vkk}
the possibility that it corresponds to a primordial magnetic field produced at the QCD phase transition, with a characteristic scale of $k_* \sim 10$ has been proposed.\cite{Brandenburg:2021tmp,Neronov:2020qrl}
Moreover, if the magnetic field is non-helical, the amplitude and coherence scales at
recombination are compatible with those recently proposed to relieve the Hubble tension.\cite{Jedamzik:2020krr}
This is a topic currently under investigation by the author and collaborators.
Note, however, that the NANOGrav observations cannot be confirmed yet
to correspond to a GW signal.

\begin{figure}
\begin{minipage}{0.5\linewidth}
\includegraphics[width=\linewidth]{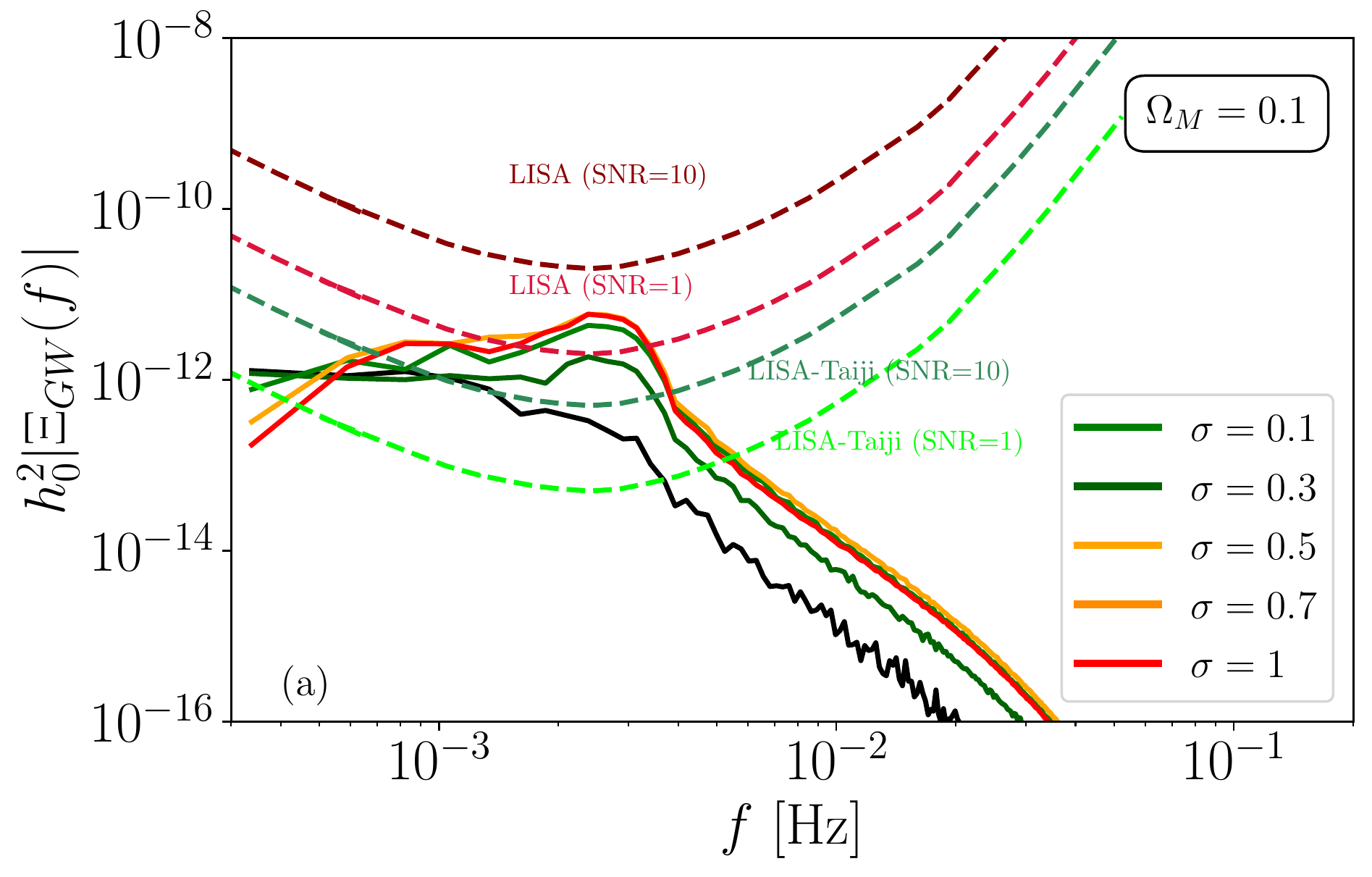}
\end{minipage}
\hfill
\begin{minipage}{0.5\linewidth}
\centerline{\includegraphics[width=\linewidth]{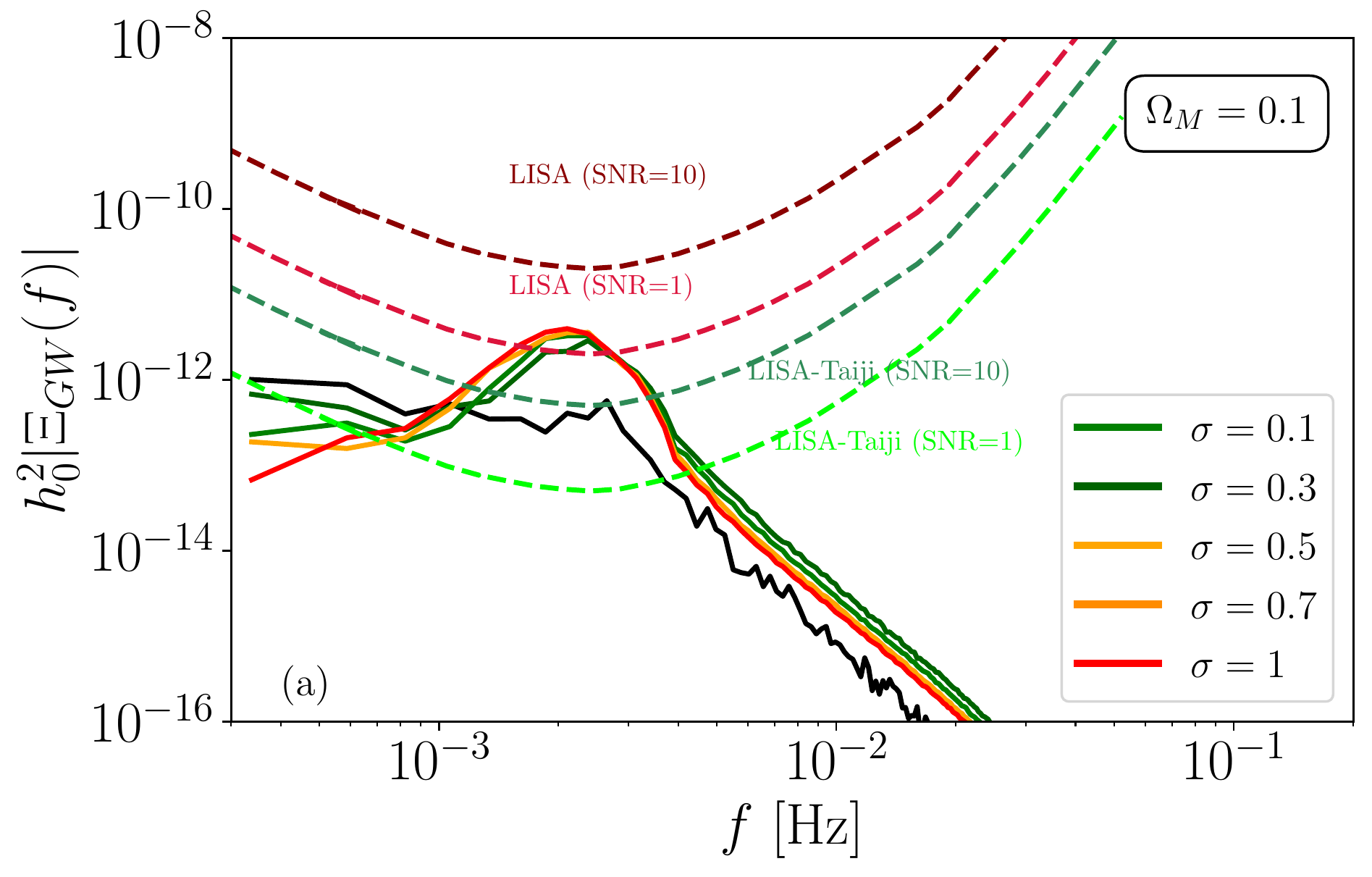}}
\end{minipage}
\hfill
\caption[]{Helical GW spectrum produced by primordial kinetic fields (left) and magnetic fields (right)
from the electroweak phase transition using
the numerical results of Kahniashvili et al. 2021\,\cite{Kahniashvili:2020jgm} compared to the
power law sensitivities computed from the proposed methods to detect a polarized signal with LISA\,\cite{Domcke:2019zls} and by combining LISA and Taiji.\cite{Orlando:2020oko}
A range of fractional helicities of the primordial field $\varepsilon=2\sigma/(1+\sigma^2)$  and
different types of turbulence sourcing have been considered (unpublished work in progress).}
\label{fig:radish}
\end{figure}

\section{Conclusions}

The new era of gravitational astronomy is at its beginning stage, and it has already led to great 
discoveries, challenging our understanding on astrophysics.
Future detections will keep providing us with new information of our Universe,
and with new challenges, even at scales that are not accessible to traditional astronomy.
One of the many cosmological sources of GWs is MHD turbulence.
The numerical implementation of a GW module within the {\sc Pencil Code} 
has been shown to be very useful as a tool to compute GW  signals produced at the
early universe, which combined with analytical calculations, can shed light into
our understanding of the spectral shape of cosmological GW backgrounds.
It is particularly useful to address the turbulence dynamics of the velocity and/or magnetic fields,
which required a case-specific turbulence modelling in previous analytical calculations.
The GW spectral shape highly depends on the mechanism of turbulence, and
the range of possible characteristic turbulent scales and sourcing amplitudes
can be explored using numerical simulations.
The resulting GW signals produced by MHD turbulence at the EWPT are potentially detectable by 
future space-based GW detector LISA, and a novel $\Omega_{\rm GW} \sim f$ spectrum that extends 
from the Hubble scale to the frequency corresponding to twice the characteristic sourcing scale,
has been shown to appear while the turbulence source is acting, in time scales related to the
turbulence scale.
If the recent PTA observations are proven to be produced by a background of gravitational waves, this 
would be consistent with the presence of primordial magnetic fields at the QCD phase
transition with a characteristic scale near the horizon, and could also be compatible with the presence 
of primordial magnetic fields at recombination.
The latter have been proposed to be able to relieve the Hubble tension.
The detection of such GW signals would provide {\bf \em clean} information on the turbulence sources
of the GW spectrum, e.g., primordial magnetic fields (that could explain the current observations of 
magnetic fields at cosmic void scales) or velocity fields induced by the expansion of first-order phase 
transition bubbles, among other sources.
The potential detection of the circular polarization of GWs by LISA (or by combining the results from 
LISA and an additional space-based GW detector, e.g., TianQin) can provide us with information of 
parity-violation processes in the early-universe, which can be related to, for example,
the presence of helicity in magnetic fields at later times and to electroweak baryogenesis.

\section*{Acknowledgments}

I would like to acknowledge my collaborators, in alphabetical order, A. Brandenburg, C. Caprini,
G. Gogoberidze, T. Kahniashvili, A. Kosowsky, S. Mandal, A. Neronov, D. Semikoz, A. Tevzadze,
and T. Vachaspati for their fruitful contributions to the present work.
Support through the French National Research Agency (ANR) project MMUniverse (ANR-19-CE31-0020) and the Shota Rustaveli National Science Fundation of Georgia (grant FR/18-1462)
are gratefully acknowledged.
I thank the {\em Rencontres de Moriond} organizers for providing a great environment
during the remote meeting and to give me the chance to contribute to the meeting.

\end{document}